\documentclass[cits]{PoS}

\title{AGN emission processes of NGC~4945 in the X-rays and $\gamma$-rays}

\ShortTitle{AGN emission processes of NGC~4945 in the X-Rays and $\gamma$-Rays}

\author{\speaker{Marie-Luise Menzel}\thanks{Now at the Helmholtz-Zentrum Dresden Rossendorf and the University of Technology Dresden, Germany.}, Volker Beckmann, and Fabio Mattana\\
                 Fran\c{c}ois Arago Centre, APC, Universit\'e Paris Diderot, CNRS/IN2P3, CEA/Irfu, Observatoire de Paris, Sorbonne Paris Cit\'e, 10 rue A. Domon et L. Duquet, 
75205 Paris Cedex 13, France\\
        E-mail: \email{ml.menzel@t-online.de}}

\abstract{NGC~4945 has an outstanding role among the Seyfert 2 active galatic nuclei (AGN) because it is one of the few non-blazars which have been detected in the $\gamma$-rays. Here, we analyse the high energy spectrum using {\it Suzaku}, {\it INTEGRAL} and {\it Fermi} data. We reconstruct the spectral energy distribution in the soft X-ray to $\gamma$-ray domain in order to provide a better understanding of the processes in the AGN. We present two models to fit the high-energy data. 

The first model assumes that the $\gamma$-ray emission originates from one single non-thermal component, e.g. a shock-induced pion decay caused by the starburst processes in the host galaxy, or by interaction with cosmic rays. The second model describes the high-energy spectrum by two independent components: a thermal inverse Compton process of photons in the non-beamed AGN and a non-thermal emission of the $\gamma$-rays. These components are represented by an absorbed cut-off power law for the thermal component in the X-ray energy range and a simple power law for the non-thermal component in the $\gamma$-rays. For the thermal process, we obtain a photon index of $\Gamma=1.6$, a cut-off energy of $E_\mathrm{cut}\approx150$\,keV and a hydrogen column density of $N_\mathrm{H}^\mathrm{hard}\approx 6\times10^{24}$ cm$^{-2}$. The non-thermal process has a photon index of $\Gamma=2.0$ and a flux of $F_{0.1-100 \rm \, GeV}=1.4\times10^{-11} \rm \, erg \,  cm^{-2} \, s^{-1}$. The spectral energy distribution gives a total unabsorbed 
flux of $F_{2 \rm\, keV-100 \rm \, GeV}\approx 5\times10^{-10}  \rm \, erg \, cm^{-2} \, s^{-1}$  and a luminosity of $L_{2 \rm\, keV-100 \rm \, GeV} \approx 9 \times10^{41} \rm \, erg \, s^{-1}$ at a distance of 3.7~Mpc.
It appears more reasonable that the $\gamma$-ray emission is independent from the AGN and could be caused e.g. by  shock processes in the starburst regions of the host galaxy. 
}

\FullConference{The Extreme and Variable High Energy Sky\\
September 19-23, 2011\\
Chia Laguna (Cagliari), Italy}

\begin{document}

\section{Introduction}
NGC~4945 is a nearby galaxy ($z=0.0019$ or 3.7 Mpc) 
in the southern hemisphere at $RA = 13^{\rm h} \, 05^{\rm m} \, 27^{\rm s}$ 
and $DEC = -49^\circ \, 28^{\rm m} \, 06^{\rm s}$ (J2000.0) which is expected to belong to the Centaurus galaxy group \cite{Hesser84}. It is an almost edge-on spiral galaxy (inclination angle $i=78^{\circ}$ \cite{Ott01}) of type SBcd or SABcd. 

NGC~4945 exhibits a prominent dust lane crossing its plane and hosts a bright and compact core. 
It is an example of a galaxy with a composite nature of an Seyfert 2 active galactic nuclei (AGN) and a star-forming region \cite{Chou07}. It shows AGN X-ray emission as well as a strong infrared component originating from starburst activity. NGC~4945 is the brightest AGN above 100 keV and one of the best studied Compton-thick sources with an intrinsic column density of the order of $N_{\rm H} = 5 \times 10^{24} \rm \, cm^{-2}$ \cite{Iwasawa93}. 
From the definition of radio loudness as $R^* = f_{5 \rm GHz}/f_{\rm optical}$ NGC~4945 classifies as a radio quiet object, with an optical core brightness of $\sim 9.5^{\rm mag}$ and a radio flux of $\sim 3 \rm \, Jy$, resulting in $R^* \sim 4.5$. Recently, Teng et al. \cite{Teng11} applied a different measure for the radio loudness, based on the radio to X-ray ratio. Here, NGC~4945 would qualify as slightly radio loud when compared to other Seyfert galaxies (by using the X-ray or hard X-ray flux). 

At high energies NGC~4945 is one of the few non-blazars which have been detected in the $\gamma$-rays. This detection came as a surprise, because non-beamed AGN and starburst galaxies were not expected to produce significant emission at these high energies. This also motivated our study to investigate the spectral energy distribution of NGC~4945 in greater detail.
Here we use unpublished hard X-ray data of NGC~4945 from the {\it INTEGRAL} and {\it Suzaku} X-ray satellites, and published results from {\it Fermi}/LAT. 
We analyze the combined X-ray and gamma-ray spectra in order to test different models to explain the processes driving the overall emission. In considering the star formation 
in the host galaxy and the processes in the AGN core, the physical connection between the emission processes of the X-ray to the $\gamma$-ray spectrum is investigated.
On one hand, thermal inverse Compton is assumed to dominate the spectrum of Seyfert galaxies in the X-ray domain, where the contribution of the starburst is expected to be faint. On the other hand, both the starbust activity (e.g., via collisionless shocks) and the AGN can contribute to the $\gamma$-ray yield. It is important to disentangle these components in order to clarify the physical mechanisms dominating in AGN.

\section{Data analysis}

The spectrum of NGC~4945 which was analyzed in this work has been observed by instruments on-board {\it Suzaku}, {\it INTEGRAL} and {\it Fermi}. In Table~\ref{mission01}, we give the characteristics of the observations.

\begin{table} [h!t] \centering
 \begin{tabular} [c] {|l|l|l|l|l |lll}\hline
mission & instrument & net exposure time & energy range used \\
\hline\hline
{\it Suzaku} & XIS (X0, X2, X3)&285.3 ks 
&4.5 - 8.5 keV\\
{\it Suzaku }& HXD-PIN & 84.9 ks 
&15 - 40 keV\\\hline

{\it INTEGRAL} & JEM-X & 152.5 ks 
&10 - 35 keV \\
{\it INTEGRAL} & ISGRI & 67.2 ks 
&40 - 250 keV\\
{\it INTEGRAL} & SPI & 530 ks 
&25 - 70 keV\\\hline

{\it Fermi} & LAT & 36.8 Ms 
&150 MeV - 2.5 GeV\\\hline
\end{tabular}
\caption{Observation log of the data used in this study.}\label{mission01}
\end{table}

From {\it Suzaku}, we used data of XIS (X-ray Imaging Spectrometer) and HXD-PIN (Hard X-ray Detector) based on the observation of 15 January 2006. 
The XIS total spectrum has been derived by summing up the XIS-0, XIS-2 and XIS-3 spectra. The background-corrected spectra have been extracted from a 90'' radius region centered on the source and data in the 4.5 -- 8.5 keV band have been used. The spectrum from the non-imaging HXD detector have been background and dead time corrected following the standard analysis. The cosmic X-ray background (CXB) is approximately 5\% of the NXB (non X-ray background) component and was estimated with a high-energy cut-off power law by using the HXD-PIN response for diffuse emission 
in \texttt{XSPEC}. HXD-PIN data have been extracted in the 15 -- 40 keV band and the spectrum has been rebinned to have at least 300~counts per channel.
In addition, we used data from the instruments SPI, ISGRI and JEM-X onboard the {\it INTEGRAL} satellite. Here we summed up the available data since the start of the mission in late 2002. A significant spectrum was extracted from SPI data in the 25 -- 70 keV band. IBIS/ISGRI data showed some calibration problems at low energies and thus we used data only within 40 -- 350 keV, and JEM-X spectra have been extracted from 10 -- 35 keV.
 
We used {\it Fermi} Large Area Telescope (LAT) observations covering August 2008 until end of March 2011. The data have been extracted from a 3.3$^{\circ}$ location around the AGN in order to avoid contamination by other sources in the region. The dead time corrected data have been rebinned to 5 energy bins in the range 150 MeV -- 2.5 GeV.
Because of the low-statistics in source counts, we applied a simple power law 
in the \texttt{gtlike} analysis of the {\it Fermi}/LAT data. The background has been modeled by an extragalactic component with a flux of $F_\mathrm{extragal}= 3.1\times 10^{-4} \, \rm ph \, cm^{-2} \, s^{-1}$ and a Galactic background of $F_\mathrm{Gal}= 4.8 \times 10^{-4} \rm \, ph \, cm^{-2} \, s^{-1}$. As result, we obtain a photon index of $\Gamma=2.3 \pm 0.1$, a $TS$-value of 92.9 (about $9.6 \sigma$) and a flux of $F_{(> 100 \rm MeV)} = (2.2 \pm 0.5) \times 10^{-8} \rm \, ph \, cm^{-2} \, s^{-1}$. Our fit parameters are consistent with the results derived by Lenain et al. \cite{Lenain10}.  

\section{Spectral energy distribution}

In order to interpret the multiwavelength spectrum in the 10 keV -- 2.5 GeV range we focus on two models to explain the X-ray to $\gamma$-ray spectrum: {\it a)} a non-thermal model or {\it b)} a combined thermal and non-thermal model. The analysis was performed using the \texttt{XSPEC} spectral fitting package version 12.5.1.

First, we will assume that the X-ray to $\gamma$-ray emission originates from one single non-thermal component. This component could be explained e.g. by mutiple shocks caused by the starburst processes in the host galaxy, or by interaction with cosmic rays. A simple absorbed power law model (\texttt{wabs*powerlaw}) gives  a photon index of $\Gamma=2.1\pm0.1$, a hydrogen column density of $N_{\rm H} = (6.8 \pm 0.6) \times 10^{24} \rm \, cm^{-2}$ with a goodness of fit of $\chi_\nu^2 = 1.06$ for 66 degrees of freedom (d.o.f.). Adding an exponential cut-off to the data (\texttt{wabs*cutoffpl}) gives a photon index of $\Gamma= 1.5 \pm 0.4$, with an exponential cut-off of $E_\mathrm{cut} < 478 \rm \, keV$. This model is a poorer representation of the data (with $\chi_\nu^2 = 1.2$ for 65 d.o.f.), and in addition it is not able to reproduce the LAT data points. 

The high-energy spectrum can also be described by two independent components. The first component is assumed to result from the photons from the thermal inverse Compton processes in the non-beamed AGN. 
The X-ray component is fitted by a cut-off power law (\texttt{wabs*cutoffpl}) described by the photon index $\Gamma_1$, the cut-off energy $E_\mathrm{cut}$ and the hydrogen column density $N_\mathrm{H}$. The second component, responsible for the emission at $\gamma$-rays is non-thermal and thus fit by an additional simple power law described with photon index $\Gamma_2$. The fit results show that \texttt{wabs} \texttt{*cutoffpl+powerlaw} is the best model for the X-ray spectrum ($\chi^2_{\nu}=1.01$, 63 d.o.f.). Assuming an absorbed cut-off power law leads to a photon index of $\Gamma_1=1.6\pm0.4$, an upper limit for the cut-off energy of $E_\mathrm{cut} < 230 \rm \, keV$, and a hydrogen column density of $N_\mathrm{H} \sim (6 \pm 2) \times 10^{24} \rm \, cm^{-2}$. The non-thermal component can be fit with a slightly steeper photon index of $\Gamma_2= 2.0^{+0.2}_{-0.5}$. Adding a reflection component to the model (e.g. \texttt{pexrav}) does not lead to an improved fit.

For the complete SED (4.5 keV -- 2.5 GeV), we apply the model \texttt{wabs\,*(gausslines\,+} \texttt{powerlaw\,+\,mekal)\,+(\,wabs\,*\,cutoffpl)}\,+\,{\texttt{powerlaw}. The first expression describes the absorbed soft X-ray energy range with a complex of five 
iron and nickel lines (\texttt{gausslines} where every single line is modeled by a Gaussian function), a photon scattering component (\texttt{powerlaw}) and thermal starburst component (\texttt{mekal}). 
We fixed the line energies of the fluorescence emissions (Fe\,{\sc i} K$\alpha_1$, K$\alpha_2$, K$\beta$; Fe\,{\sc xxv} K$\alpha$; Ni\,{\sc i} K$\alpha$) to their expected positions according to the redshift of NGC~4945. Their positions are consistent with the values found when fitting the XIS spectrum only (Table~\ref{compcomp}). The second expression is fitting the transmitted photons with a heavily absorbed exponential cut-off power law (\texttt{wabs*cutoffpl}) and has the same photon index as the soft power law component. The third expression describes the non-thermal shock component in the $\gamma$-rays as explained above.
We can reproduce the spectrum of the AGN in the X-rays, the starburst (at soft X-rays), and the non-thermal shock component in the $\gamma$-rays, achieving a fit statistics of $\chi^2_\nu=1.16$ (98 d.o.f). The combined data fits are shown in Figure \ref{plots} and the fit results are summarised in Table~\ref{compcomp}, including the iron and nickel lines detectable in the {\it Suzaku}/XIS spectrum. We obtain a total absorption corrected flux of $F_{2 \rm\, keV-100 \rm \, GeV} \sim 5 \times 10^{-10} \rm \, erg \, cm^{-2} \, s^{-1}$ and a luminosity of $L_{2 \rm\, keV-100 \rm \, GeV} \sim 9 \times 10^{41} \rm \,erg \, s^{-1}$. 





\begin{table}\centering
\begin{minipage}[t]{0.35\textwidth}
\begin{tabular} {|l|c|}\hline
parameter & value \\\hline\hline 
$\Gamma_1$	&	$1.6 \pm 0.4$ 	\\
$N_{\rm H}^\mathrm{soft}$ ($10^{22} \rm \, cm^{-2}$) 	&	$7$  (fixed)	\\
$kT$ (keV)	&	$3 {+3 \atop -1}$ 	\\\hline\hline

$E_\mathrm{cut}$ (keV)	&	$150$ (fixed)	\\
$N_{\rm H}^\mathrm{hard}$ ($10^{23} \rm \, cm^{-2}$) 	&	$59 {+7 \atop -5}$	\\\hline\hline

$\Gamma_2$	&	$2.0 {+0.2 \atop -0.5}$\\ \hline\hline

$\chi^2_\nu$ (d.o.f.)	&	1.16 (98)	\\ \hline
\end{tabular}
\end{minipage}
\begin{minipage}[t]{0.35\textwidth}
\begin{tabular}{|l|c|c|}\hline
line & energy [keV] & $EW$ [eV]\\\hline\hline
Fe\,{\sc i} K$\alpha_1$ & $6.40\pm0.01$ & $400 {+100 \atop -50}$ \\
Fe\,{\sc i} K$\alpha_2$ & $6.25 {+0.15 \atop -0.06}$ &	$80 {+40 \atop -70}$\\
Fe\,{\sc xxv} K$\alpha$ & $6.68 \pm 0.02$ & $140 {+30 \atop -20}$ \\
Fe\,{\sc i} K$\beta$    & $7.04 {+0.02 \atop -0.03}$ & $130 \pm 30$\\
Ni\,{\sc i} K$\alpha$ & 7.47 (fix) &  $70 \pm 50$ \\\hline
\end{tabular}
\end{minipage}
\caption{Parameters of the fit. Errors have been determined on the X-ray spectrum without the {\it Fermi}/LAT data. For the iron line measurements, only the {\it Suzaku}/XIS data have been used.}\label{compcomp}
\end{table}

\begin{figure}[h!c] \centering
\begin{minipage}[c]{0.49\textwidth}
\includegraphics[width=\textwidth]{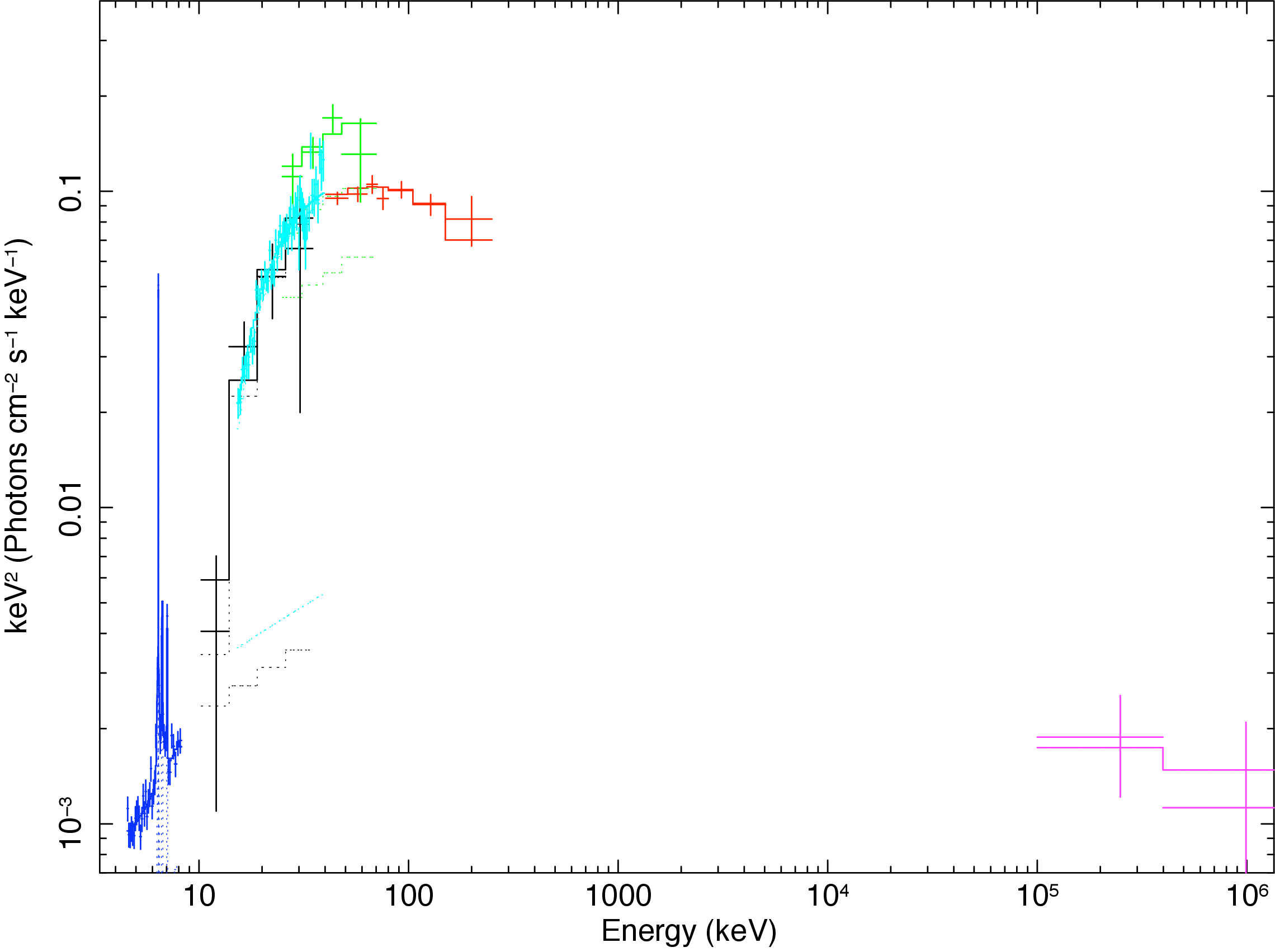}
\end{minipage}
\begin{minipage}[c]{0.49\textwidth}
\includegraphics[width=\textwidth]{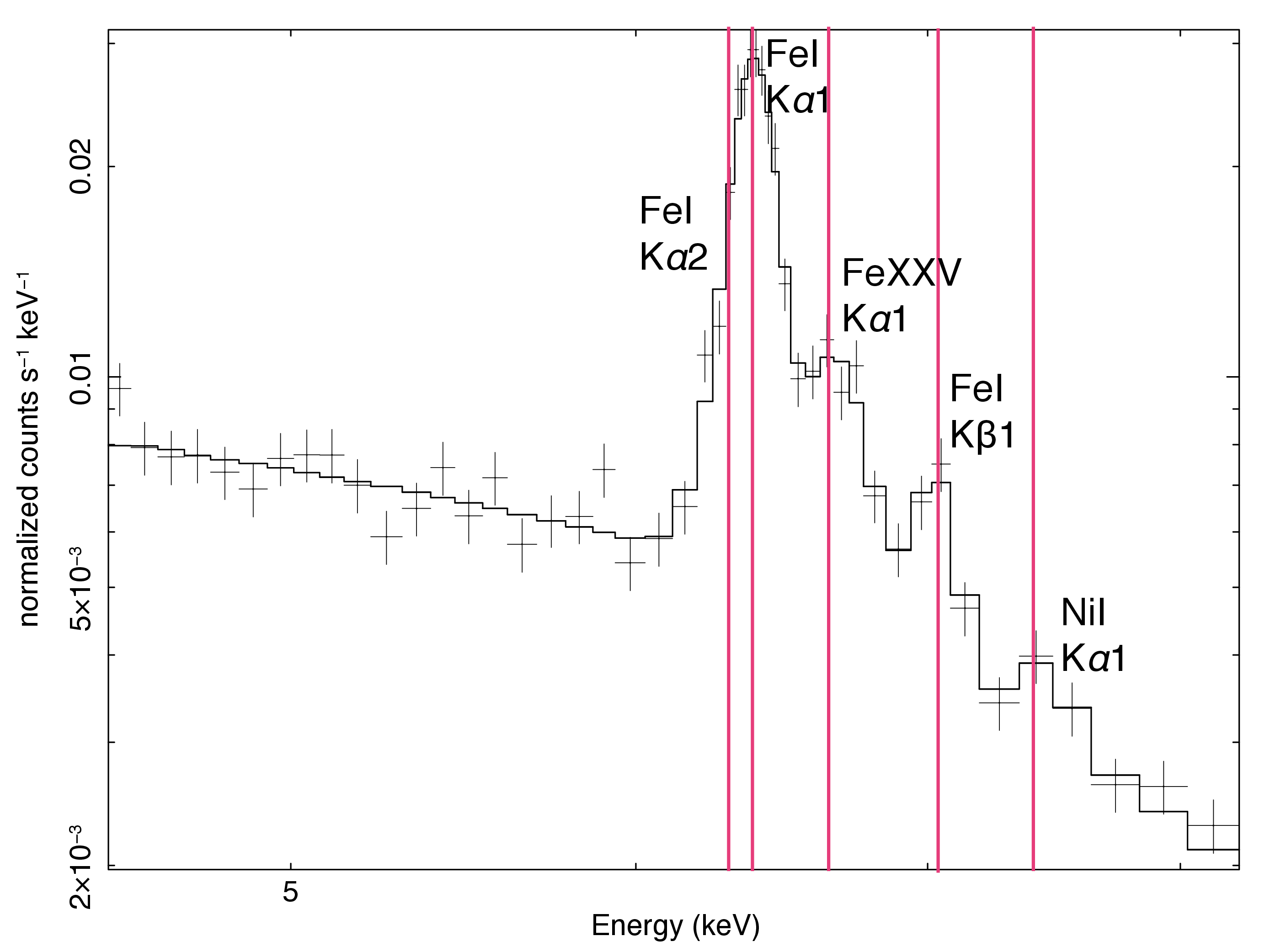}
\end{minipage}
\caption{Left panel: unfolded spectrum $E^2f(E)$, Right panel: Soft X-ray spectrum with Gaussian lines }\label{plots}
\end{figure}

\section{Discussion}

The model we applied to the combined data allows a representation of the spectral energy distribution over the whole $4.5 - 10^{6} \rm \, keV$ energy range we studied. For the X-ray range, we applied the models for the AGN and starburst at the same time. We obtain following fitting results: a photon index of $\Gamma=1.6$, a cut-off energy of $E_\mathrm{cut} \sim 150$ keV, an absorber in the soft X-rays (surrounding star burst component)  with $N_{\rm H}^\mathrm{soft}= 7 \times 10^{22} \rm \,  cm^{-2}$ and $N_{\rm H}^\mathrm{hard} \sim 6 \times 10^{24} \rm \, cm^{-2}$ for the AGN core emission itself. For the starburst we obtain a temperature of $kT = 3 \rm \,  keV$. These results are consistent with previous analysis of the {\it Suzaku} data alone \cite{Itoh08}, who used exactly the same data set but an earlier software and calibration version.
In the $\gamma$-ray domain the detection significance of NGC~4945 is still rather low. With the data we obtained of the AGN, we can reproduce the results of Lenain et al. \cite{Lenain10} which was based on an earlier {\it Fermi}/LAT data set. 

It is difficult to constrain the model for the combined $\gamma$-ray and hard X-ray data, because of the wide gap in the $\sim 200 \rm \, keV$ to 100 MeV range.
We applied two different models to explain the $\gamma$-ray emission. In the first scenario, we assume that the $\gamma$-ray and hard X-ray data are correlated and can be modeled by one non-thermal model. Here, the $\gamma$-ray and the X-rays have to originate from the same inverse Compton process. In order to reach the $\gamma$--ray domain, in this case one would need to assume that NGC~4945 hosts a jet, which gives rise to this non-thermal inverse Compton component.
In the second model, we assume a thermal inverse Compton component for the hard X-ray spectrum of the AGN and a second non-thermal component, represented by the simple power law. Because of the low significance and relatively small bandwidth of the detection of this component, we can only assume that the $\gamma$--rays are caused by the star burst in this case, for example by shock processes. Similar emission has been seen by {\it Fermi}/LAT in the star forming galaxies NGC~253 and M~82. One can think of this as a superposition of many Eta carina like systems, where a massive stellar wind hits the interstellar medium and thus produces strong shocks which give rise to the $\gamma$--ray emission. Alternatively, cosmic ray interactions in the interstellar medium of the host galaxy can be responsible \cite{Lenain10}.

We obtain the best fit results with the second scenario, i.e. two independent components.
Thus, it is more likely that the high-energy spectrum consists of a combination of a thermal (inverse Compton) and non-thermal component. NGC~4945 hosts no blazar, there is no jet observed in the source, and there is no strong radio emission which would be an indication for one single non-thermal component responsible for the hard X-ray spectrum. Also the presence of the strong iron and nickel fluorescence lines confirms the thermal and typical Seyfert-type AGN emission. The photons of the AGN disc are most likely the main source for the ionization and fluorescence processes in the AGN surrounding matter. 

For the spectrum in the $2 - 10^8$ keV energy range, we obtain a total absorption corrected flux of $F_{2 \rm\, keV- 100 \rm \, GeV} \sim 5 \times 10^{-10} \rm \,  erg \, cm^{-2} \, s^{-1}$ and a luminosity of $L_{2 \rm\, keV-100 \rm \, GeV} \sim 9 \times 10^{41} \rm \, erg \, s^{-1}$. This flux and luminosity are similar to the value in the energy range of 2 -- 300 keV. Thus, the X-ray energy range remains a good proxy for the total high-energy emission of AGNs, even if we consider the more complex case with $\gamma$--ray emission as observed in NGC~4945.

Assuming that the high-energy emission is about half of the total bolometric emission, we obtain an Eddington ratio of $\lambda \sim 0.005$ for NGC~4945. Seyfert~2 galaxies normally tend to have a lower average Eddington ratio than Seyfert~1 galaxies ($\langle \lambda_\mathrm{Sy2} \rangle =0.02$ compared to $\langle \lambda_\mathrm{Sy1} \rangle = 0.06$, \cite{Beckmann09}). 
Other authors even find lower Eddington ratios, e.g. $\lambda=0.006$ for 87 AGN at $z<1.25$ as presented by Simmons et al. \cite{Simmons11}. On the other hand, low mass black holes accrete on average at higher Eddington ratios \cite{Gallo10}. Peculiar Seyfert galaxies can have Eddington ratios close to the Eddington limit ($\lambda=1$), like MCG--5--23--16 \cite{Beckmann08}, or even display super-Eddington accretion with $\lambda>1$ \cite{Woo02}. 
In summary, it appears that the Seyfert core in NGC~4945 is operating at a rather "standard" or low Eddington rate. The complexity of the X-ray spectrum with the transmitted, reflected, and star-burst component, is mainly due to the high signal-to-noise one can obtain for a near-by Seyfert galaxy. In the case of NGC~4945 the $\gamma$--ray emission is not linked to the central engine, thus Seyfert galaxies remain to be no $\gamma$-ray emitters, unless one considers the star burst in the host galaxy. The pre-requisite for $\gamma$-ray emission from the AGN remains to be the existence of a powerful jet, like in the case of the $\gamma$-ray bright source Cen~A \cite{Beckmann11}.

\end{document}